# EFFECTS OF MAGNETIC AND AHARANOV-BOHM (AB) FIELDS ON THE ENERGY SPECTRA OF THE YUKAWA POTENTIAL

**COLLINS EDET***

University of Port Hartcourt, Faculty of Science, Department of Physics, Rivers State, Nigeria

*Corresponding Author's Email Address: collinsokonedet@gmail.com

## ABSTRACT

*In this article, the Yukawa potential is scrutinized taking into consideration the effects of magnetic and AB flux fields within the non-relativistic regime using the factorization method. The energy equation and wave function of the system are obtained in close form. We find that the all-encompassing effects result in a strongly attractive system and consequently there is a significant upward shift in the bound state energy of the system. We also find that to achieve a low-energy medium for the Yukawa potential, weak magnetic field is required, however the AB flux field can be used as a controller. The magnetic and AB fields eliminates the degeneracy from the spectra. From our findings, it could be concluded that to manipulate the energy spectra of this system, the AB-flux and magnetic field will do so greatly. The results from this study can be applied in condensed matter physics, atomic and molecular physics.*

## 1. INTRODUCTION

In non-relativistic quantum mechanics (QM), the Schrödinger equation (SE) is a very relevant equation. This is due to the fact that it is the equation that aids our understanding of the dynamics of a particle in the microscopic environment. Solving the SE with a particular interaction potential is of huge relevance. Furthermore, several authors have noted that the eigensolutions (eigenvalues and eigen-function) of the SE incorporates significant information vis-à-vis the potential models [1-10]. Many works with regard to solutions of the SE for several of potential models have been considered by several researchers [5-10]. One of such potentials is the Yukawa Potential (YP). This model was first proposed by Hideki Yukawa [11]. This potential emerged when there was a loophole in the description of the James Chadwick's atomic model, which was made up of protons and neutrons packed inside of a small nucleus with a radius of $10^{-4} m$. This inspired him to propose the screened coulomb or Yukawa potential which includes an exponential decay term $\left(e^{-\alpha r}\right)$ and an electromagnetic term $\left(\dfrac{1}{r}\right)$. This model was an extension and a general case of the long-range Coulombic interaction potential [12]. YP has vast and increasing applications in other fields of physics such as; high energy physics, atomic and solid state physics, amongst others [13]. The YP can be written as [11,12,13]:

$$V(r) = -\frac{V_1 e^{-\delta r}}{r} \qquad (1)$$

where $r$ is the inter-particle distance, $V_1$ is the potential parameter and $\delta$ is the screening parameter which characterize the range of the interaction[11,12,13].

In this research, we are interested in providing answers to the following questions; what happens to the energy spectra of this model in the presence of the all-inclusive effect of magnetic and Aharanov-Bohm (AB) fields? What happens when there is a solitary effect? This questions motivated us to carry out this study. Lately, many researchers have focused their attention on studies that considers the influence of magnetic and AB fields on the energy spectra of myriads of QM systems [14-19].

In the present work, our goal is to solve the SE with the Yukawa potential (YP) model in the presence of magnetic and AB flux fields using the factorization method. We will discuss the effects of the fields on the energy spectra of the system.

The paper is organized as follows. In section 2, we solve of the 2D Schrödinger equation with the Yukawa Potential (YP) and vector potential $\vec{A}(r)$ under the combined effects of magnetic and AB flux fields. In section 3, we discuss the effects of the fields on the behavior of the energy spectra of the YP. Finally, a brief concluding remark is given in section 4.

## 2. THEORY AND SOLUTIONS

The Hamiltonian operator of a particle that is charged and confined to move in the region of YP under the collective impact of AB flux and magnetic fields can be written in cylindrical coordinates. Thus, the Schrödinger equation for this consideration is written as follows [14, 18, 19];

$$\left[\frac{1}{2\mu}\left(i\hbar\vec{\nabla} - \frac{e}{c}\vec{A}\right)^2 - \frac{V_1 e^{-\delta r}}{r}\right]\psi(r,\varphi) = E_{nm}\psi(r,\varphi), \tag{2}$$

where $E_{nm}$ denotes the energy level, $\mu$ is the effective mass of the system, the vector potential which is denoted by "$\vec{A}$" can be written as a superposition of two terms $\vec{A} = \vec{A}_1 + \vec{A}_2$ having the azimuthal components [14] and external magnetic field with $\vec{\nabla} \times \vec{A}_1 = \vec{B}, \vec{\nabla} \times \vec{A}_2 = 0$, where $\vec{B}$ is the magnetic field. $\vec{A}_1 = \frac{\vec{B}e^{-\delta r}\hat{\varphi}}{(1-e^{-\delta r})}$ and $\vec{A}_2 = \frac{\phi_{AB}}{2\pi r}\hat{\varphi}$ represents the additional magnetic flux $\phi_{AB}$ created by a solenoid with $\vec{\nabla} \cdot \vec{A}_2 = 0$. The vector potential in full is written in a simple form as [14, 18, 19];

$$\vec{A} = \left(0, \frac{\vec{B}e^{-\delta r}}{(1-e^{-\delta r})} + \frac{\phi_{AB}}{2\pi r}, 0\right) \tag{3}$$

Let us take a wave function in the cylindrical coordinates as $\psi(r,\varphi) = \frac{1}{\sqrt{2\pi r}}e^{im\varphi}R_{nm}(r)$, where $m$ denotes the magnetic quantum number. Inserting this wave function and the vector potential into Eq. (2) we arrive at the following radial second-order differential equation:

$$R''_{nm}(r) + \frac{2\mu}{\hbar^2}\left[E_{nm} - V_{eff}(r)\right]R_{nm}(r) = 0 \tag{4}$$

where $V_{eff}(r)$ is the effective potential defined as follows;

$$V_{eff}(r,\omega_c,\xi) = -\frac{V_1 e^{-\delta r}}{r} + \hbar\omega_c(m+\xi)\frac{e^{-\delta r}}{(1-e^{-\delta r})r} + \left(\frac{\mu\omega_c^2}{2}\right)\frac{e^{-2\delta r}}{(1-e^{-\delta r})^2} + \frac{\hbar^2}{2\mu}\left[\frac{(m+\xi)^2 - \frac{1}{4}}{r^2}\right] \tag{5}$$

where $\xi = \frac{\phi_{AB}}{\phi_0}$ is an integer with the flux quantum $\phi_0 = \frac{hc}{e}$ and $\omega_c = \frac{e\vec{B}}{\mu c}$ denotes the cyclotron frequency. Eq. (4) is a complicated differential equation that cannot be solved easily due to the presence of centrifugal term. Therefore, we employ the Greene and Aldrich approximation scheme [3, 4, 6] to bypass the centrifugal term. This approximation is given as;

$$\frac{1}{r^2} = \frac{\delta^2}{(1-e^{-\delta r})^2} \tag{6}$$

We point out here that this approximation is only valid for small values of the screening parameter $\delta$. Inserting Eqs. (6) into Eq. (4) and introducing a new variable $s = e^{-\delta r}$ allows us to obtain

$$\frac{d^2 R_{nm}(s)}{ds^2} + \frac{1}{s}\frac{dR_{nm}(s)}{ds} + \frac{1}{s^2(1-s)^2}\left[-(\varepsilon_{nm}+\beta_0+\beta_2)s^2 + (2\varepsilon_{nm}+\beta_0-\beta_1)s - (\varepsilon_{nm}+\eta)\right]R_{nm}(s) = 0 \quad (7)$$

For Mathematical simplicity, we have introduced the following dimensionless notations;

$$-\varepsilon_{nm} = \frac{2\mu E_{nm}}{\hbar^2 \delta^2}, \beta_0 = \frac{2\mu V_1}{\hbar^2 \delta}, \beta_1 = \frac{2\mu\omega_c}{\hbar\delta}(m+\xi), \beta_2 = \frac{\mu^2\omega_c^2}{\hbar^2\delta^2} \text{ and } \eta = (m+\xi)^2 - \frac{1}{4} \quad (8)$$

In order to solve eq. (7), we have to transform the differential in equation (7) into a form solvable by any of the existing standard mathematical technique. Hence, we take the radial wave function of the form

$$R_{nm}(s) = s^\lambda (1-s)^\nu f_{nm}(s) \quad (9)$$

where: $\lambda = \sqrt{\varepsilon_{nm}+\eta}; \quad \nu = \frac{1}{2} + \sqrt{\frac{1}{4} + \beta_2 + \beta_1 + \eta} \quad (10)$

On substitution of Eq. (9) into Eq. (7), we obtain the following hypergeometric equation:

$$s(1-s)f''_{nm}(s) + \left[(2\lambda+1) - (2\lambda+2\nu+1)s\right]f'_{nm}(s) - \left[(\lambda+\nu)^2 - \left(\sqrt{\varepsilon_{nm}+\beta_0+\beta_2}\right)^2\right]f_{nm}(s) = 0 \quad (11)$$

From eq. (11), by considering the finiteness of the solutions, the quantum condition is given by

$$(\lambda+\nu) - \left(\sqrt{\varepsilon_{nm}+\beta_0+\beta_2}\right) = -n \qquad n = 0,1,2... \quad (12)$$

which in turn transforms into the energy eigenvalue equation as follows;

$$E_{nm} = \frac{\hbar^2\delta^2\eta}{2\mu} - \frac{\hbar^2\delta^2}{8\mu}\left[\frac{\frac{2\mu V_1}{\hbar^2\delta} + \frac{\mu^2\omega_c^2}{\hbar^2\delta^2} - \eta - \left(n+\frac{1}{2}+\sqrt{\frac{\mu^2\omega_c^2}{\hbar^2\delta^2} + \frac{2\mu\omega_c}{\hbar\delta}(m+\xi) + (m+\xi)^2}\right)^2}{\left(n+\frac{1}{2}+\sqrt{\frac{\mu^2\omega_c^2}{\hbar^2\delta^2} + \frac{2\mu\omega_c}{\hbar\delta}(m+\xi) + (m+\xi)^2}\right)}\right]^2 \quad (14)$$

Consequently, the wave function is obtained as follows;

$$R_{nm}(s) = s^{\sqrt{\varepsilon_{nm}+\eta}}(1-s)^{\frac{1}{2}+\sqrt{\frac{1}{4}+\beta_2+\beta_1+\eta}} \,_2F_1\left((\lambda+\nu)+\left(\sqrt{\varepsilon_{nm}+\beta_0+\beta_2}\right), (\lambda+\nu)-\left(\sqrt{\varepsilon_{nm}+\beta_0+\beta_2}\right), 2\lambda+1; s\right) \quad (15)$$

## 3. RESULTS AND DISCUSSION

Table 1 shows the numerical energy values for the Yukawa potential under the influence of AB flux and magnetic fields with various values of magnetic quantum numbers. We observe that when both fields are absent $(\vec{B} = \xi = 0)$, there exist degeneracy in the energy spectra. By introducing only magnetic field $(\vec{B} \neq 0, \xi = 0)$ to the system, the energy eigenvalues is increased and takes away the degeneracy as well. Nevertheless, as the strength of the magnetic field is raised so is the energy. This suggests that the energy values of the Yukawa can be altered or regulated to a highest level by the application a strong magnetic field. The application of the AB field only $(\vec{B} = 0, \xi \neq 0)$, raises the energy values and degeneracies are eliminated. The energy spectra become more negative and the system becomes strongly attractive as the quantum number $n$ increases for fixed $m$. The combined effect $(\vec{B} \neq 0, \xi \neq 0)$ of both fields is robust and therefore, there is an upward shift in the bound state energy of the system. The combined effect completely eliminates the degeneracy. The complete effects shows that the system is strongly attractive while the localizations of quantum levels change and the eigenvalues increase. Also, the combined effect of the fields is strong and consequently, there is a significant upward shift in the bound state energy of the system. The effect of the magnetic field is seen to be stronger than the combined effect.

**Table 1**: The energy levels of the Yukawa potential (YP) with various $n$ and $m$ quantum states for $D=2$ in the presence and absence of external magnetic field, $\vec{B}$ and AB flux fields, $\xi$. With the following fitting parameters; $\hbar = \mu = e = c = 1; \delta = 0.005; V_1 = 2$

| $m$ | $n$ | $\omega_c = \xi = 0$ | $\omega_c = 5, \xi = 0$ | $\omega_c = 0, \xi = 5$ | $\omega_c = \xi = 5$ |
|---|---|---|---|---|---|
| 0 | 0 | -8.0000031 | -0.0000032 | -0.0570279 | -0.000016 |
|  | 1 | -0.8844531 | -0.0000182 | -0.0394309 | -0.000155 |
|  | 2 | -0.3152211 | -0.0000581 | -0.0284107 | -0.000319 |
|  | 3 | -0.1584072 | -0.0001229 | -0.0210664 | -0.000507 |
| -1 | 0 | -0.8822253 | $-7.49 \times 10^{-7}$ | -0.0898797 | -0.000013 |
|  | 1 | -0.3144151 | $9.25 \times 10^{-7}$ | -0.0585341 | -0.000128 |
|  | 2 | -0.1579929 | -0.0000057 | -0.0405253 | -0.000267 |
|  | 3 | -0.0936389 | -0.0000457 | -0.0292467 | -0.000430 |
| 1 | 0 | -0.8822253 | $-5.75 \times 10^{-6}$ | -0.0381096 | -0.000018 |
|  | 1 | -0.3144151 | -0.0000457 | -0.0274011 | -0.000182 |
|  | 2 | -0.1579929 | -0.0001104 | -0.0202652 | -0.000370 |
|  | 3 | -0.0936389 | -0.0001999 | -0.0152829 | -0.000583 |

**4. CONLUSION**

In this research article, the Yukawa or screened Coulomb potential is analyzed with the influence of magnetic and AB flux fields. The Hamiltonian operator containing the external fields and the potential model is transformed into a second-order differential equation. We solve this differential equation via the factorization method to acquire the energy equation and wave function of the system. The effect of the fields on the energy spectra of the system is closely examined. It was found out that the magnetic and AB fields remove degeneracy.